\documentclass[10pt,letterpaper,twocolumn]{article} 

\usepackage{ol2ori}
\usepackage{amsmath}
\usepackage{graphicx}

\begin{document}

\twocolumn[ 

\title{Spectral control of broadband light through random media by wavefront shaping}

\author{Eran Small$^1$, Ori Katz$^1$, Yefeng Guan$^{1,2}$, and Yaron Silberberg$^1$}
\address{$^1$ Department of Physics of Complex Systems, The Weizmann Institute of Science, Rehovot 76100, Israel\\
$^2$ State Key Laboratory of Optoelectronic Materials and Technologies, Sun Yat-sen University, Guangzhou 510275, China}

\begin{abstract}
A random medium can serve as a controllable arbitrary spectral
filter with spectral resolution determined by the inverse of the
interaction time of the light in the medium. We use wavefront
shaping to implement an arbitrary spectral response at a particular
point in the scattered field. We experimentally demonstrate this
technique by selecting either a narrow band or dual bands with a
width of 5.5nm each.

\end{abstract}


] 

\noindent

The propagation of light in disordered media results in scattering, which
is often a major limitation in many optical applications, ranging from microscopy
to astronomy. However, optical wave propagation through disordered media can be controlled and even inverted using wavefront-shaping \cite{MoskOL} or phase-conjugation \cite{Yaqoob}, a result that has been extensively studied in time-reversal techniques in acoustics and radio-frequency electromagnetic waves \cite{Fink,FinkSci}. Perhaps
surprisingly, random scattering can be harnessed rather than fought against to surpass the diffraction-limit \cite{FinkSci,MoskNP}, to obtain temporal control by means of spatial manipulation only \cite{Katz,Aulbach}, or,
vice versa, to achieve spatial control by temporal manipulation
\cite{Mccabe,FinkPRL}. Related to these results, it was long ago suggested by Freund
\cite{Freund} that a characterized random medium can be used as an arbitrary
optical component. Indeed, it was recently shown that a random medium can be used as a lens \cite{MoskOL,MoskNP,Gigan1,Gigan2}, a mirror\cite{Wide}, a spectral-polarimetric analyzer \cite{Dogariu}, a pulse shaping device \cite{Katz,Aulbach} and an arbitrary waveplate
\cite{Yefeng}. Here we demonstrate the transformation of the random
medium into an arbitrary spectral filter by wavefront shaping.

When an ultrashort pulse is scattered in a random medium, it
undergoes both spatial and temporal distortions. As result of the spatio-temporal coupling in the random medium, both the spatial and temporal distortions can be corrected by manipulating the spatial shape of the incident wavefront \cite{Katz,Aulbach}. Since the scattering medium couples the spatial and temporal degrees
of freedom, it also couples the spatial and spectral degrees of
freedom. Here we exploit the same coupling to control the spectral
profile of a broadband source (not necessarily an ultrashort pulsed source) focused through the random medium by
manipulating its wavefront. We show that this technique can be used
to generate arbitrary optical filters at a controlled focus.

\begin{figure}[htB]
\centerline{\includegraphics[width=8.5cm]{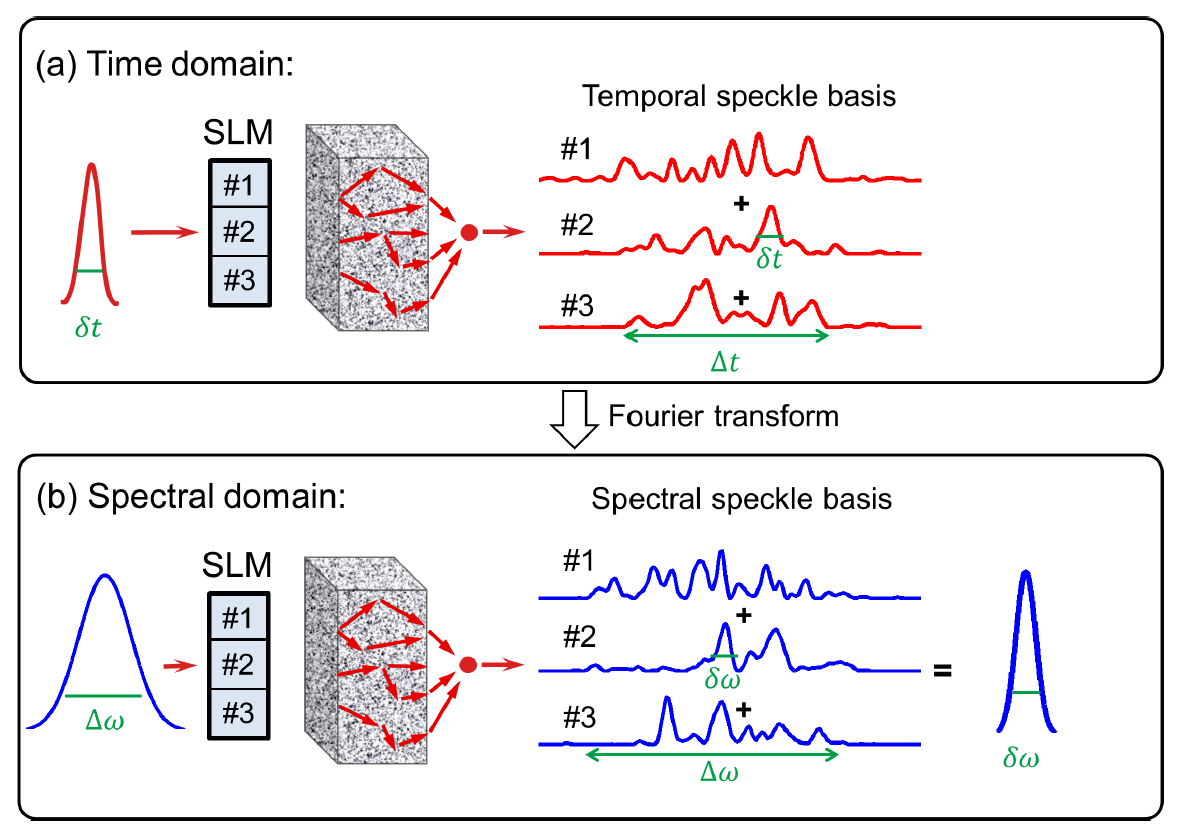}}
\caption{\label{speckles} {\footnotesize Scattering in a random
medium couples each SLM pixel to a different temporal (and spectral)
pattern when measured at a particular output point, thus forming a
random effective basis for temporal/spectral control by the SLM. In the
temporal domain (a) an ultrashort pulse with duration of $\delta t$
at the input is scattered into different temporal patterns
characterized by the same minimal feature size $\delta t$ and a
temporal span $\Delta t$ determined by the interaction time of the
light inside the medium. In the spectral domain (b) a broadband
input with a bandwidth of $\Delta\omega\propto\frac{1}{\delta t}$ is
scattered into different spectral patterns, characterized by a
minimal feature size $\delta\omega\propto\frac{1}{\Delta t}$ and
spanning the full $\Delta\omega$ of the source. A phase only SLM at
the input can thus control the coherent interference of these random
spectral patterns, generating an arbitrary optimized spectral pattern at
the output.}}
\end{figure}

To implement these ideas, a phase-only spatial light modulator (SLM) is used to
shape the incident wavefront. The light passing through each of the
SLM pixels generates a different temporal speckle pattern after
propagating through the scattering medium (Fig.
\ref{speckles}a). These different random temporal patterns form an
effective temporal basis for temporal control. Consequently, as each temporal  patten is related to a particular spectral pattern by Fourier relation, it provides an effective spectral basis (Fig.
\ref{speckles}b) which can be
used to achieve spectral control, with a resolution determined by the
interaction time of the light in the medium.

Schematic sketches of these temporal and spectral speckle patterns
are presented in Fig.\ref{speckles}. The temporal patterns plotted
in Fig.\ref{speckles}(a) are characterized by two properties: the
shortest temporal feature, or {\lq\lq}temporal speckle" of
characteristic duration $\delta t$, and the total temporal span
$\Delta t$. $\delta t$ is determined by the bandwidth of the
broadband laser source $\delta t \propto \frac{1}{\Delta \omega}$,
while the temporal span $\Delta t$ is determined by the interaction
time of the light in the scattering medium, and by optical path
length differences when using an off-axis configuration (when
considering the spectrum of a spot which is not on the optical axis)
\cite{Small}. The spectral characteristics of the speckle patterns
presented in Fig.\ref{speckles}(b) are related to the temporal
patterns by Fourier transform: The spectral span $\Delta\omega$ is
simply the bandwidth of the broadband source which is inversely
proportional to the temporal speckle width
$\Delta\omega\propto1/\delta t$, while the narrowest spectral
feature ({\lq\lq}spectral speckle") $\delta\omega$ is dictated by
the total temporal span $\delta\omega\propto1/\Delta t$. In terms of
wavelength the minimal speckle size is given by:

\begin{equation}
\delta\lambda=\frac{\lambda_0^2}{c\Delta t} \label{dw}
\end{equation}

where $\lambda_0$ is the central wavelength of the broadband source
and $c$ is the speed of light. The spectral speckle width,
$\delta\lambda$, dictates the spectral resolution of the random
medium as a spectral filter or as a pulse shaping device \cite{Katz}.
While for a standard grating-based spectrometer the resolution is
determined by the number of grating lines illuminated by the beam,
here the resolution is determined by $\delta\lambda$ which is a
property of the scattering medium and the geometry. It is not hard
to see that, perhaps counter intuitively, a more scattering sample
yields higher spectral resolution.

As in other utilizations of wavefront-shaping\cite{MoskOL}, the
control is obtained by optimizing the phases of the SLM elements to
minimize a cost function, which in this case is defined as the
difference from a desired spectral function at the spatial measurement
position. We note that only a small fraction of the incident energy is actually
involved in the process.

\begin{figure}[htB]
\centerline{\includegraphics[width=8.2cm]{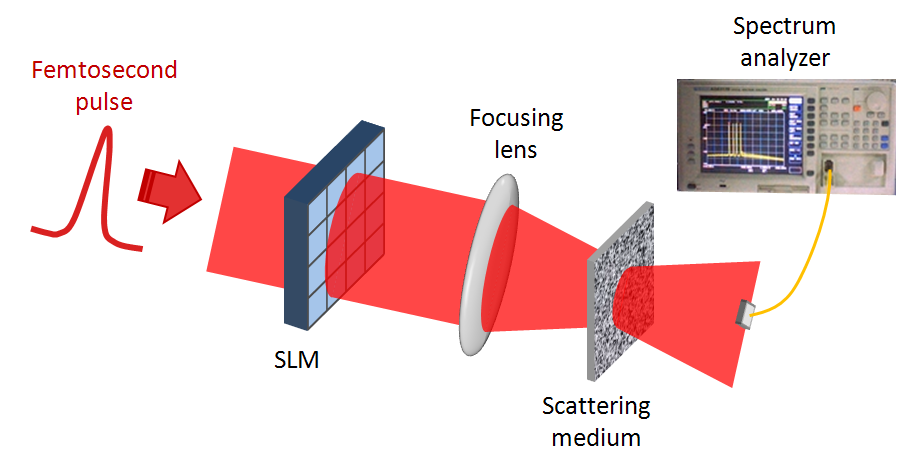}}
\caption{\label{setup} {\footnotesize Experimental setup for
spectral control of a broadband source through a random medium by
wavefront shaping. An SLM controls the wavefront of the incident
spatially-coherent broadband field (not necessarily an ultrashort pulse). The wavefront shaped field is
focused on the scattering medium with a lens. The scattered light at
a particular spatial position behind the medium is measured by an
optical spectrum analyzer, which generates a feedback signal for
optimizing the incident wavefront to attain the
desired spectral pattern.}}
\end{figure}

The experimental setup for spectral control through a random medium by wavefront-shaping is shown in Fig.\ref{setup}. The simple
experimental setup consists of a spatially-coherent broadband laser source (a
femtosecond laser Idesta Octavius-1G), a phase only 2-dimensional
SLM (Hamamatsu model X10468-01) , a focusing lens, a scattering
medium (Polycarbonate diffuser) and a fiber-coupled spectrum
analyzer (Ando model AQ-6315A). We use the SLM to shape the
wavefront of the incident field which is then focused with the lens
on the scattering medium (note that the lens is not required to
achieve spectral control, and is only used in order to increase the
measured intensities). The scattered light at a single point behind the medium is coupled to the optical spectrum analyzer by a
multimode fiber which is placed $~1cm$ from the scattering medium. For optimal spectral contrast, a single spatial speckle should be coupled to the fiber. The experiment takes place by optimizing the SLM phase-pattern using a genetic algorithm \cite{Genetic}
to yield the desired spectral shape on the optical spectrum analyzer.

It is important to notice that since the feedback signal is the
shape of the spectrum measured by the spectrum analyzer, the
experiment is insensitive to the spectral phases. Therefore, the
requirements from the source used for spectral control are only
spatial coherence and broad spectrum, and it does not have to be a
temporally coherent laser source (such as the ultrashort laser
source used in our experiments). This means that any spatially
coherent broadband source such as super-luminous diodes,
super-continuum sources, or spatially-filtered thermal sources
could be used as well.

\begin{figure}[htB]
\centerline{\includegraphics[width=8.4cm]{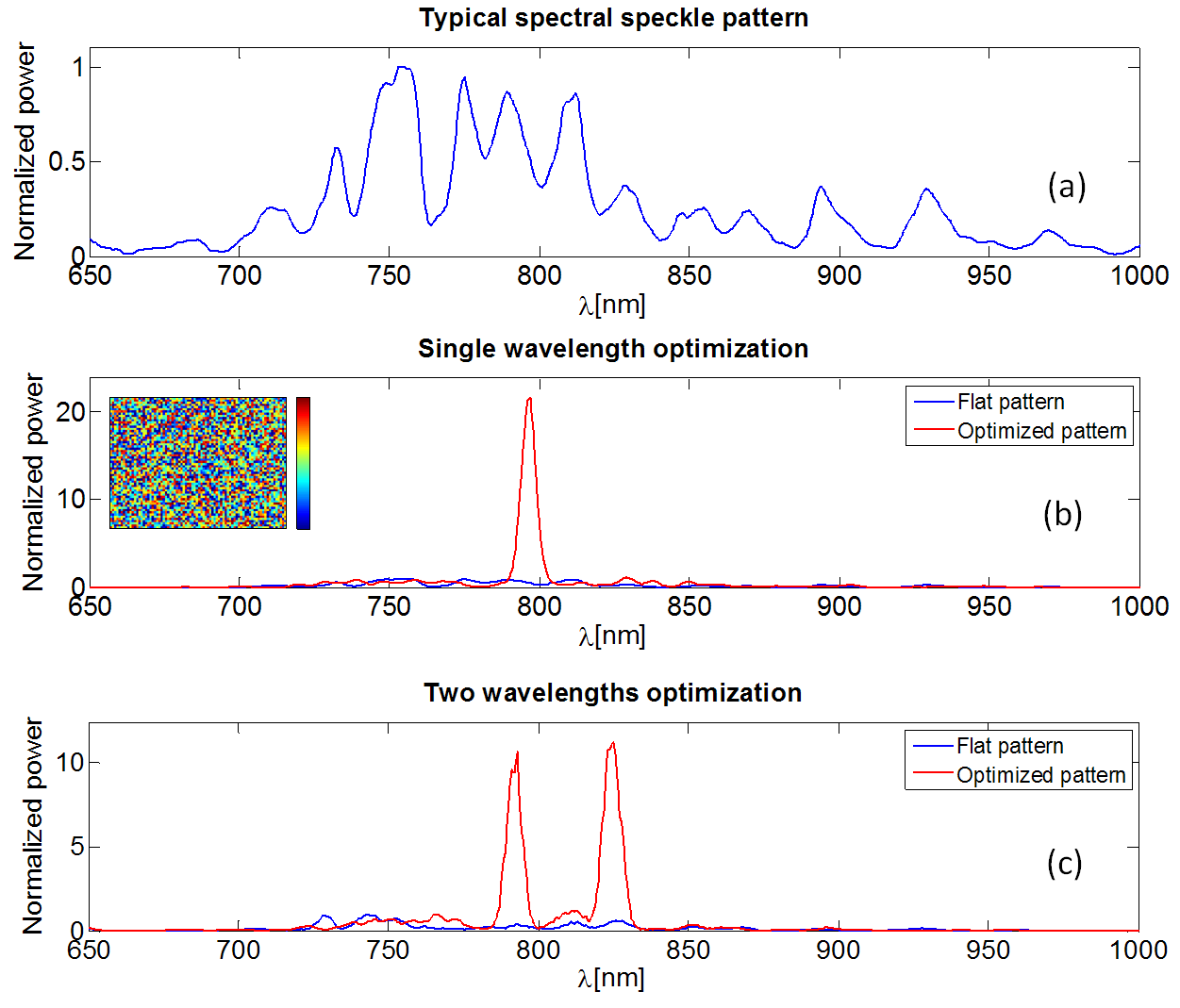}}
\caption{\label{results} {\footnotesize Experimental results: (a)
unoptimized spectrum of the scattered broadband source at the measurement point. (b)
Optimized (red line) and unoptimized (blue line) spectra for a
narrow bandpass filter optimization (inset: SLM optimized phase
pattern, the maximum phase is $2.3\pi$). (c) Optimized (red line)
and unoptimized (blue line) spectra for a dual bandpass filter
optimization. All spectral powers are normalized to the peak power
of the unoptimized spectrum.}}
\end{figure}

To demonstrate the narrowest spectral filter this method allows, we
simply choose one spectral feature in the measured scattered
spectrum presented in Fig.\ref{results}(a) and maximize its power.
The result of this optimization is shown in Fig.\ref{results}(b)
where the blue line shows the initial spectrum and the red line
shows the resulting spectrum after optimization. The inset of
Fig.\ref{results}(b) is the SLM patten that yields the optimal
result. The full width half maximum (FWHM) of the optimized spectrum
is $5.5nm$ (spectrometer resolution is $0.1nm$) and the intensity
enhancement comparing to the average unoptimized power is
approximately $50$. From this FWHM and with the help of Eq.\ref{dw},
the scattering interaction time can be estimated to be $\Delta t=400
fs$.

In order to demonstrate more complex filters, we present the results
of a dual bandpass filter in Fig.\ref{results}(c). Here we
choose two different spectral speckles and maximize their mutual
power. This procedure results in two separated narrow spectral transmission bands. The
FWHM of each one of the two lobes is the same as for the single
lobe. However, similar to spatial focusing experiments\cite{MoskOL}, the enhancement of each of the two lobes compared to
the background is half of the enhancement achieved for a single
band. This contrast is determined by the number of degrees of control\cite{MoskOL}, and is the limit for the complexity of the controlled filter.

In summary, we have shown that random scattering can be exploited for spatial and spectral control of a broadband source inside/through a random medium by wavefront shaping. This technique can generate an
arbitrary spectral filter at a specific spatial position. One can envisage using this tool for
controlled spectroscopy inside a random-medium\cite{spectroscopy}, or for spectrally resolved imaging or information transmission \cite{Gigan2} through a random medium. Different spatial output points can be simultaneously optimized to different spectral
filters using a single phase pattern. This can enable
transformation of a random medium into a grating \cite{Freund} or an optimized spectrum analyzer.
Interestingly, in analogy to the spatial "memory effect" in thin scattering media \cite{memeffect}, performing our experiment using a thin scattering medium and an off-axis geometry will result in spectrally scaling the single-point optimized spectral function when the distance of the measurement point from the optical axis is varied, using the same SLM phase-pattern. This is the result of the correlation between speckle patterns at different wavelengths in a thin scattering medium (e.g. in surface scattering) \cite{Small}. Optimizing a single wavelength at
a single point would thus lead to dispersing the full spectrum wavelength-by-wavelength in the
radial direction.

E.S. is supported by the Adams Fellowship Program of the Israel
Academy of Sciences and Humanities. This work was supported also by
MINERVA and the ERC grant QUAMI.


\begin{thebibliography}{99}

\bibitem{MoskOL} I.M. Vellekoop, and A.P. Mosk, Opt. Lett. {\bf 32}, 2309-2311 (2007).

\bibitem{Yaqoob} Z. Yaqoob, D. Psaltis, M.S. Feld, and C. Yang, Nat. Photon. {\bf 2}, 110-115 (2008).

\bibitem{Fink} M. Fink, Phys. Today {\bf 50}, 34-40 (1997).

\bibitem{FinkSci} G. Lerosey, J. de Rosny, A. Tourin, and M. Fink, Science {\bf 315}, 1120-1122 (2007).

\bibitem{MoskNP} I.M. Vellekoop, A. Lagendijk, and A.P. Mosk, Nat. Phot. {\bf 4}, 320-322 (2010).

\bibitem{Katz} O. Katz, E. Small, Y. Bromberg, and Y. Silberberg, Nat. Phot. {\bf 5}, 372-377 (2011).

\bibitem{Aulbach} J. Aulbach, B. Gjonaj, P. M. Johnson, A. P. Mosk, and A. Lagendijk, Phys. Rev. Lett. {\bf 106}, 103901 (2011).

\bibitem{Mccabe} D. J. McCabe, A. Tajalli, D.R. Austin, P. Bondareff, I. A. Walmsley, S. Gigan, and B. Chatel, Nat. Comm. {\bf 2}, 447 (2011).

\bibitem{FinkPRL} F. Lemoult, G. Lerosey, J. de Rosny, M. Fink, Phys. Rev. Lett. {\bf 103}, 173902 (2009).

\bibitem{Freund} I. Freund, Physica A: Statistical Mechanics and its Applications  {\bf 168}, 49-65 (1990).

\bibitem{Gigan1} S. M. Popoff, G. Lerosey, R. Carminati, M. Fink, A. C. Boccara, and S. Gigan, Phys Rev Lett {\bf 104}, 100601 (2010).

\bibitem{Gigan2} S. Popoff, G. Lerosey, M. Fink, A. C. Boccara, and S. Gigan, doi:10 1038/ncomms1078 (2010).


\bibitem{Wide} O. Katz, E. Small, and Y. Silberberg, arXiv 1202.2078 (2012).

\bibitem{Dogariu} T. Kohlgraf-Owens, and A. Dogariu, Opt. Lett. {\bf 35}, 2236-2238 (2010).

\bibitem{Yefeng} Y. Guan, O. Katz, E. Small, J. Zhou and Y. Silberberg, in preparation.

\bibitem{Small} E. Small, O. Katz, and Y. Silberberg, Opt. Express {\bf 20} 5189-5195 (2012).

\bibitem{Genetic} D. B. Conkey, A. N. Brown, A. M. Caravaca-Aguirre, and R. Piestun, Opt. Express {\bf 20} 4840-4849 (2012).

\bibitem{spectroscopy} T. Svensson, E. Adolfsson, M. Lewander, C. T. Xu, and S. Svanberg, Phys. Rev. Lett. {\bf 107}, 143901 (2011).

\bibitem{memeffect} I. Freund, M.Rosenbluh, and S. Feng, Phys. Rev. Lett. {\bf 61}, 2328-2331 (1988).

\end{thebibliography}
\end{document}